\documentclass[aps,prb,groupedaddress,amsmath,amssymb,twocolumn]{revtex4}

\usepackage{amssymb}
\usepackage{amsmath}
\usepackage{graphicx}
\usepackage{subfigure}
\usepackage{textcomp}
\usepackage{color}
\usepackage{amsfonts}
\usepackage{bbold}
\usepackage{dsfont}
\usepackage{epsfig}
\usepackage{hyperref}

\begin{document}
\title{Floquet dynamics in two-dimensional semi-Dirac semimetals and three-dimensional Dirac semimetals}

\author{Awadhesh Narayan}
\email{awadhesh@illinois.edu}
\affiliation{School of Physics, Trinity College, Dublin 2, Ireland,}
\affiliation{Department of Physics, University of Illinois at Urbana-Champaign, Urbana, Illinois, USA.}

\date{\today}

\begin{abstract}
 We study the photoresponse of two-dimensional semi-Dirac semimetals and three-dimensional Dirac semimetals to off-resonant circularly polarized light. For two-dimensional semi-Dirac semimetals we find that incident light does not open a gap in the spectrum, in contrast to the case of purely linear dispersion. For the three-dimensional case we find that applying a circularly polarized light, one can tune from a trivial insulator to three-dimensional Dirac semimetal with an inverted gap. We propose and show that application of light leads to an intriguing platform for generation, motion and merging of three-dimensional Dirac points.
\end{abstract}

\maketitle

\section{Introduction}

Since the rise of graphene, a two-dimensional Dirac semimetal, Dirac fermions have been under intense focus in the last decade.~\cite{review-geim} A semi-Dirac semimetal dispersion has been proposed by Dietl \textit{et al.} in honeycomb lattice in a magnetic field,~\cite{montambaux-semidirac} and by Pardo and Pickett in TiO$_2$/VO$_2$ nanostructures by means of first principles calculations.~\cite{pickett-semidirac0} This unexpected zero gap state shows a linear dispersion along one direction, while exhibiting quadratically dispersing bands along the orthogonal direction, thus the name: semi-Dirac semimetal. Interestingly, such a semi-Dirac state has also been suggested in the Hofstadter spectrum,~\cite{montambaux-hofstadter} as well as in a dielectric photonic crystal.~\cite{wu-photonic}

On the other hand, the three-dimensional analogs of graphene, namely three-dimensional Dirac semimetals were proposed theoretically.~\cite{kane-semimetal,fang-na3bi,fang-cd3as2} These semimetals exhibit linearly dispersing bands along all the three momentum directions and the Fermi surface consists of discrete Dirac points. Recently a number of experiments have found their signatures by means of angle-resolved photoemission experiments,~\cite{chen-na3bi,hasan-na3bi,chen-cd3as2,hasan-cd3as2,cava-cd3as2} as well as by using landau level spectroscopy and quasiparticle interference.~\cite{yazdani-cd3as2}

Interaction of light with Dirac fermions is also an avenue that has seen extensive recent activity. Application of light has been predicted to generate two-dimensional Floquet topological insulators,~\cite{aoki-floquet,inoue-floquet,lindner-floquet} while illuminating graphene and silicene with circularly polarized light is expected to generate a Haldane-like mass gap.~\cite{kitagawa-floquet,auerbach-graphene,ezawa-silicene,piskunow-graphene,zhai-epitaxial} An exciting new development has been the observation of Floquet-Bloch bands and light-induced band gap on the surface of topological insulators. By applying intense ultrashort midinfrared pulses Wang \textit{et al.} were able to observe polarization dependent band gaps, using time- and angle-resolved photoemission technique.~\cite{gedik-exp} This opens up the possibility to realize optical control over topological phenomena.

In light of these promising developments, in this contribution, we study the effect of off-resonant circularly polarized light on two-dimensional semi-Dirac semimetals, as well as three-dimensional Dirac semimetals. Our analytical results show that for semi-Dirac semimetals circularly polarized light does not lead to any gap in the band structure in contrast to the well-studied Dirac systems graphene and silicene. For three-dimensional case we find that illumination with light allows one to tune from a trivial insulator to a Dirac semimetal with an inverted band gap. Curiously, at the critical zero gap point during this topological transition, the band structure of the system is described by a three-dimensional semi-Dirac semimetal Hamiltonian. We show that illumination of light leads to exciting prospects for generating, manipulating, as well as annihilating Dirac points. 

\section{Results}

\subsection{Two-dimensional semi-Dirac semimetal} 

\begin{figure*}[t]
\begin{center}
  \includegraphics[scale=0.60]{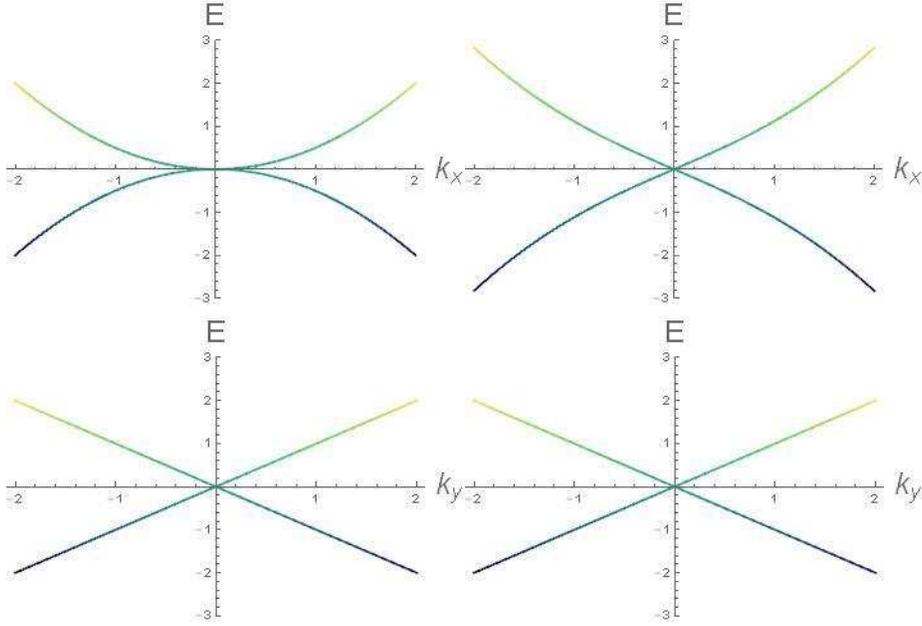}
  \caption{(Color online) The band structure for semi-Dirac semimetal along quadratically ($k_{x}$) and linearly ($k_{y}$) dispersing directions (top and bottom panels, respectively), with $\mathcal{A}=0$ (left) and $\mathcal{A}=1.0$ (right). Photoillumination does not open a gap in the spectrum but allows changing the effective group velocity. Here we set $v_{F}=1.0,m=1.0,\omega=1.0$.}  \label{2d_semidirac_semimetal}
\end{center}
\end{figure*}

Let us begin with the analysis of semi-Dirac semimetal in two-dimensions under illumination by off-resonant light. The low-energy Hamiltonian for a semi-Dirac semimetal can be written as~\cite{pickett-semidirac1,pickett-semidirac2}

\begin{equation}
 H=v_{F}k_{y}\sigma_{z}+\frac{k_{x}^{2}}{2m}\sigma_{x},
\end{equation}

\noindent where $v_{F}$ is the Fermi velocity along the linear dispersion direction ($k_y$), $m$ is the mass along the direction with quadratic dispersion ($k_x$), and $\sigma_{x}$ and  $\sigma_{z}$ are the $x$ and $z$ components of the triad of Pauli matrices (here and henceforth we have set $\hbar=1$). Consider illumination by circularly-polarized off-resonant light of frequency $\omega$, which generates a vector potential $\mathcal{A}(t)=\mathcal{A}(\eta \sin\omega t \hat{x}+\cos \omega t \hat{y})$. Here $\eta=\pm 1$ for right and left circularly polarized beams, respectively. The response of the electrons can be obtained by minimal coupling $k\rightarrow k+e\mathcal{A}(t)$. If we consider low order virtual processes involving only a single photon, then the time-dependent Hamiltonian can be approximated by~\cite{kitagawa-floquet}

\begin{equation}\label{ham_floquet}
 H_{\mathrm{eff}}\approx H+\frac{[H_{-1},H_{+1}]}{\omega}.
\end{equation}

\noindent Here $H_{\pm 1}=\frac{1}{T}\int_{0}^{T}H(t)e^{\pm i\omega t}$ are the Fourier components of the full time-dependent Hamiltonian, and $T=2\pi/\omega$ is the time period of the incident light. Here we note that the above is an appropriate treatment of out-of-equilibrium physics in the large-frequency limit, while the static limit is ill-defined. Computing the Fourier components of the Hamiltonian and using the commutation relations for the Pauli matrices, we obtain the commutator as $[H_{-1},H_{+1}]= -\eta\frac{e^{2}\mathcal{A}^{2}k_{x}v_{F}}{m}\sigma_{y}$. Then the effective Hamiltonian in the presence of light can be written down as

\begin{equation}
 H_{\mathrm{eff}}=v_{F}k_{y}\sigma_{z}+\frac{k_{x}^{2}}{2m}\sigma_{x}-\eta\frac{e^{2}\mathcal{A}^{2}k_{x}v_{F}}{m\omega}\sigma_{y}.
\end{equation}

\noindent In the above effective Hamiltonian, we note the peculiar form of the additional term arising from the coupling to off-resonant light involving both the momentum, $k_{x}$, along the quadratic dispersion direction, as well as the Fermi velocity, $v_{F}$, along the Dirac-like dispersion. This arises as a consequence of the unusual semi-Dirac dispersion.

\begin{figure}[b]
\begin{center}
  \includegraphics[scale=0.40]{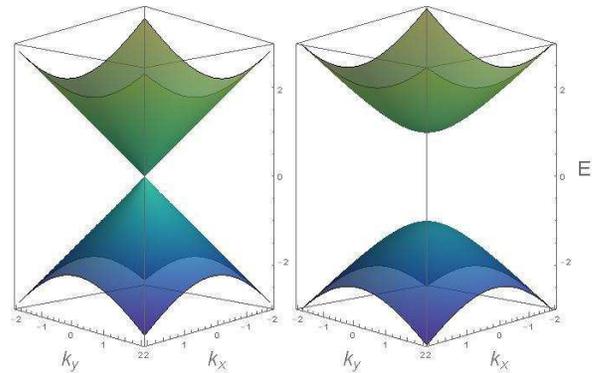}
  \caption{(Color online) Photon dressed band structure for a two-dimensional Dirac semimetal with $\mathcal{A}=0$ (left panel) and $\mathcal{A}=1.0$ (right panel). Note the gap opening for non-zero light intensity. Here we set $v_{F}=1.0,\omega=1.0$.}  \label{2d_dirac_semimetal}
\end{center}
\end{figure}

The energy eigenvalues are obtained as $E_{\pm}=\pm\frac{1}{2m}\sqrt{k_{x}^{4}+4k_{y}^{2}m^{2}v_{F}^{2}+4\eta^{2}e^{4}\mathcal{A}^{4}k_{x}^{2}v_{F}^{2}/\omega^{2}}$. Note that since $\eta^{2}=1$, the left or right-handedness of the circularly polarized light does not influence the photon dressed band structure. The photon dressed band structure for the semi-Dirac semimetal is shown in Fig.~\ref{2d_semidirac_semimetal} with $\mathcal{A}=0$ and $\mathcal{A}=1.0$, along the quadratic and linear dispersion directions. In this case light illumination does not open a gap in the spectrum, although it does change the dispersion shape along $k_{x}$. This is a result of the competition between the terms linear and quadratic in $k_{x}$, i.e., the terms proportional to $\sigma_{y}$ and $\sigma_{x}$, respectively. Increasing light intensity increases the linearity of the bands along this direction, while maintaining the linearity of bands in the orthogonal direction. In this way photoillumination allows a handle over the \textit{degree of Diracness} of the system.~\cite{goerbig-diracness}

It is illustrative to contrast the above obtained result with the case of purely linear energy-momentum relation, i.e., for a two-dimensional Dirac semimetal like graphene, with the Hamiltonian, $H=v_{F}(k_{x}\sigma_{x}+k_{y}\sigma_{y})$. In this case, the commutator reads, $[H_{-1},H_{+1}]= \eta e^{2}\mathcal{A}^{2}v_{F}^{2}\sigma_{z}$. This leads to an additional term in the Hamiltonian, which is quadratic in $v_{F}$ and serves as a mass proportional to $\sigma_{z}$. As shown in Fig.~\ref{2d_dirac_semimetal}, the consequence of this term is that the spectrum can be gapped on application of circularly polarized light. This is a Floquet counterpart of the Chern insulator, which harbors protected chiral edge modes along with a bulk energy gap and has been extensively explored recently.~\cite{kitagawa-floquet,auerbach-graphene,ezawa-silicene,piskunow-graphene,zhai-epitaxial}

\begin{figure*}[t]
\begin{center}
  \includegraphics[scale=0.50]{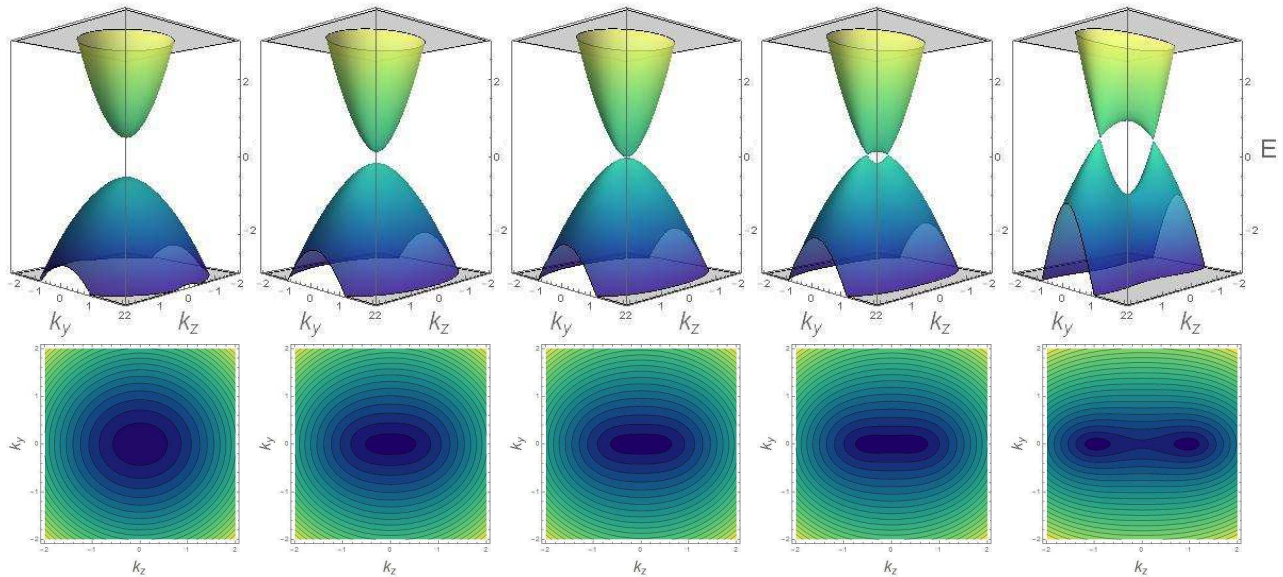}
  \caption{(Color online) Evolution of band structure (top row) and band gap (bottom row) for the three-dimensional Dirac semimetal under increasing intensity of photoillumination, $\mathcal{A}=0.0,0.6,0.7,0.8,1.2$, from left to right. Note the reduction of trivial gap with increasing $\mathcal{A}$, which becomes zero for $\mathcal{A}\approx$ 0.7. The gap reopens beyond this illumination strength, with inverted order, along with a pair of Dirac points located symmetrically along $k_{y}=0$. In the bottom row contour plots of band gap are shown with increasing $\mathcal{A}$. Darker (lighter) shades represent smaller (larger) energy gap. Here we set $c_{0}=0,c_{1}=0.5,c_{2}=0.5,m_{0}=0.5,m_{1}=-1.0,m_{2}=-1.0,A=1.0,\omega=1.0$.}  \label{3d_dirac_semimetal}
\end{center}
\end{figure*}

\subsection{Three-dimensional Dirac semimetal} 

After studying the effects of light illumination on the semi-Dirac semimetal, let us next look at its effects on a three-dimensional Dirac semimetal. A low-energy Hamiltonian for three-dimensional Dirac semimetals around the Brillouin zone center, considering terms up to quadratic in $k$, is of the form~\cite{fang-na3bi,fang-cd3as2}

\begin{equation}
 H=\epsilon_{0}(\mathbf{k})\mathbb{I}_{2\times 2} + M(\mathbf{k})\sigma_{z}+A(k_{x}\sigma_{x}-k_{y}\sigma_{y}),
\end{equation}

\noindent where $\epsilon_{0}(\mathbf{k})=c_{0}+c_{1}k_{z}^{2}+c_{2}(k_{x}^{2}+k_{y}^{2})$, $M(\mathbf{k})=m_{0}-m_{1}k_{z}^{2}-m_{2}(k_{x}^{2}+k_{y}^{2})$ and $\mathbb{I}_{2\times 2}$ is the $2\times 2$ identity matrix. The energy eigenvalues are obtained as $E_{\pm}=\epsilon_{0}(\mathbf{k})\pm\sqrt{M(\mathbf{k})^{2}+A^{2}(k_{x}^{2}+k_{y}^{2})}$. The spectrum shows Dirac crossings at $k_{x}=0,k_{y}=0,k_{z}=\pm\sqrt{m_{0}/m_{1}}$. It is interesting to note that in addition to the bulk Dirac points, the model also exhibits an inverted band order around the Brillouin zone center. This means that when confined to a two-dimensional setup, the energy spectrum shows topological surface states, in the same way as three-dimensional topological insulators exhibit two-dimensional surface states in a slab geometry.

Under the illumination by off-resonant circularly polarized light, we start by invoking the form of approximate Hamiltonian in equation~(\ref{ham_floquet}), and after some algebra, we find that in this case the commutator takes the form, $[H_{-1},H_{+1}]= -2\eta e^{2}\mathcal{A}^{2}Am_{2}(k_{x}\sigma_{x} - k_{y}\sigma_{y} ) - \eta e^{2}\mathcal{A}^{2}A^{2}\sigma_{z}$. Then, the effective Hamiltonian for a three-dimensional Dirac semimetal illuminated by circularly polarized light reads

\begin{widetext}

\begin{equation}
 H_{\mathrm{eff}}=\epsilon_{0}(\mathbf{k})\mathbb{I}_{2\times 2} + M(\mathbf{k})\sigma_{z}+A(k_{x}\sigma_{x}-k_{y}\sigma_{y}) + \eta\frac{-2e^{2}\mathcal{A}^{2}Am_{2}k_{x}}{\omega}\sigma_{x} + \eta\frac{2e^{2}\mathcal{A}^{2}Am_{2}k_{y}}{\omega}\sigma_{y} + \eta\frac{-e^{2}\mathcal{A}^{2}A^{2}}{\omega}\sigma_{z}.
\end{equation}

\end{widetext}

\noindent Note that in this case light induces terms proportional to all the three Pauli matrices, in contrast to the case of two-dimensional Dirac semimetals and semi-Dirac semimetals. This includes a term proportional to $\sigma_{z}$ which can modulate the mass, and thereby the band inversion strength. The total mass term is given by $m_{\mathrm{eff}}=M(\mathbf{k})-\eta e^{2}\mathcal{A}^{2}A^{2}/\omega$. The dependence of this term on the polarization, $\eta$, means that changing the polarization from right to left (i.e., changing $\eta$ from +1 to -1) would reverse the effect of light on the band inversion strength. Additionally, there are terms proportional to $\sigma_{x,y}$ linear in momentum, $k_{x,y}$, which serve to renormalize the velocities in the $x,y$ directions. Note, that the motion of the Dirac points and the band inversion strength is governed by the term proportional to $\sigma_{z}$. The overall effect is a possibility to create a band inversion as well as crystal symmetry protected Dirac points. To see this, let us start with a topologically trivial insulator (this is obtained by choosing $m_{0}>0$ and $\mathcal{A}=0$ in the low energy model) and switch-on the illumination. The resulting series of band structures are shown in Fig.~\ref{3d_dirac_semimetal}. As the light intensity is enhanced the band gap decreases, reaching a critical zero gap point. Beyond this intensity the gap at the Brillouin zone center reopens, with an inverted band order, and at the same time Dirac points appear away from the center. Interestingly, at the critical zero gap point the band structure is described by a three-dimensional version of the semi-Dirac Hamiltonian, ~\cite{montambaux-merging1,montambaux-merging2,montambaux-merging3} i.e., the bands disperse quadratically along $k_{z}$, while dispersing linearly along $k_{y}$ as well as $k_{x}$, as can be seen from the contour plot of the band gap in Fig.~\ref{3d_dirac_semimetal}. Our argument can also be turned around to observe merging of Dirac points, by switching the polarization of the illuminating light from right to left circularly polarized light. In this case one starts with a Dirac semimetal and with increasing intensity of the applied light, the two time-reversed cones start moving towards the center of the Brillouin zone. Subsequently, they meet at a critical point where the band inversion is zero. Further increase in light intensity would open a gap and render the material a trivial insulator. Thus, depending on the polarization of light it is possible to increase as well as reduce the band inversion strength in three-dimensional Dirac semimetals. 

Two materials Na$_3$Bi and Cd$_3$As$_2$ have been recently found to exhibit three-dimensional Dirac fermion states, along with an inverted bulk gap. Based on our proof-of-principle results, it would be interesting to expose these materials to intense circularly polarized beams to explore the suggested possibility of tuning the band inversion strength, as well as the $k$-space position of the Dirac crossings. The photoillumination method that we propose here could also be applied to lower spin-orbit cousins of these known three-dimensional Dirac semimetals Na$_3$Bi and Cd$_3$As$_2$, namely Na$_3$Sb and Cd$_3$P$_2$. This potentially mitigates the need for alloying with large spin-orbit elements to drive topological phase transitions.~\cite{sanvito-na3bi} To get an estimate for the laser intensities needed to observe the topological phase transition, consider Cd$_{3}$P$_{2}$ which has a trivial mass gap $m_{0}\approx 50$~meV at $k_{x}=k_{y}=k_{z}=0$ and group velocity of bands near Fermi level, $A\approx 5 \times 10^{5}$~m/s. Experimentally used frequencies for experiments of interest are in the thousands of terahertz regime,~\cite{gedik-exp} and we set $\omega=1000$~THz. This yields the required laser intensity as, $I\sim 10^{14}$~W/m$^{2}$. Given the recent advances in capability to manipulate the band structures of diverse set of materials using intense circularly polarized light,~\cite{gedik-exp,gedik-tmdc,wang-tmdc} we are confident that our proposal can be an exciting playground to witness generation, as well as annihilation of three-dimensional Dirac points. We note that a proposal similar to ours has been made recently to observe merging of two-dimensional Dirac points in graphene.~\cite{merging-graphene}

\section{Conclusions and summary} 

In conclusion, we have investigated Floquet dynamics in two-dimensional semi-Dirac semimetals and three-dimensional Dirac semimetals under off-resonant light. For semi-Dirac semimetals we found that illumination with circularly polarized light does not lead to band gap opening, at variance with purely linearly dispersing bands. For three-dimensional Dirac semimetals, we have shown that shining circularly polarized light leads to a modulation of the band inversion strength whose effect is opposite for right and left circularly polarized beams. Additionally, the position of the three-Dimensional Dirac points in the reciprocal space can also be tuned. Our results suggest that photoillumination of three-dimensional Dirac semimetals can be an exciting platform to investigate generation, motion and annihilation of Dirac points.\\ 

\section*{Acknowledgments} 

The work at Trinity College Dublin was supported by the Irish Research Council under the EMBARK initiative. I thank Domenico Di Sante, Taylor Hughes, Graham Kells, Silvia Picozzi, Stefano Sanvito, Smitha Vishveshwara, Lucas Wagner and Faxian Xiu for discussions and related collaborations.

\end{document}